**Title:** Spatial Clustering of Time-Series via Mixture of Autoregressions Models and Markov Random Fields

**Authors:** Hien D. Nguyen[*,1,2], Geoffrey J. McLachlan[1], Jeremy F. P. Ullmann[2], and Andrew L. Janke[2]

[1]School of Mathematics and Physics, University of Queensland, St. Lucia, Australia. [2]Centre for Advanced Imaging, University of Queensland, St. Lucia, Australia. [*]Corresponding author (h.nguyen7@uq.edu.au).

**Abstract:** Time-series data arise in many medical and biological imaging scenarios. In such images, a time-series is obtained at each of a large number of spatially-dependent data units. It is interesting to organize these data into model-based clusters. A two-stage procedure is proposed. In Stage 1, a mixture of autoregressions (MoAR) model is used to marginally cluster the data. The MoAR model is fitted using maximum marginal likelihood (MMaL) estimation via an MM (minorization–maximization) algorithm. In Stage 2, a Markov random field (MRF) model induces a spatial structure onto the Stage 1 clustering. The MRF model is fitted using maximum pseudolikelihood (MPL) estimation via an MM algorithm. Both the MMaL and MPL estimators are proved to be consistent. Numerical properties are established for both MM algorithms. A simulation study demonstrates the performance of the two-stage procedure. An application to the segmentation of a zebrafish brain calcium image is presented.

**Short Title:** Clustering of Time-Series via MoAR Models and MRFs

# 1 Introduction

Time-series data arise in many medical and biological imaging scenarios; for example, calcium imaging, electrocardiogram (ECG), electroencephalography (EEG), and functional magnetic resonance imaging (fMRI) data. In such situations, a time-series of data is recorded at each spatially dependent data unit (e.g. an electrode, a pixel, or a voxel). Upon observing these data, an interesting problem is to cluster the time-series at each data unit into similar subgroups that are spatially coherent across the image.

The clustering of time-series data is a well-studied problem. Background on the topic can be found in Liao (2005) and Esling and Agon (2012). It is clear from Esling and Agon (2012) that there are many directions of research on this problem. Given the contents of this article, we shall focus our review on the literature regarding the mixture model-based clustering of time-series data.

In Cadez et al. (2000), a mixture of Markov chains model was suggested for the clustering of data based on web browsing behavior, time-course gene expression, and red-blood cell cytograms. In Xiong and Yeung (2004), mixture of autoregressive moving-average regressions (MoARMA) models are suggested for the clustering of ECG, EEG, population, and temperature data. In Luan and Li (2003), Celeux et al. (2005), Ng et al. (2006), and Scharl et al. (2010), various specifications of mixtures of mixed-effects models are suggested for the clustering of time-course gene expression data; Wang et al. (2012) extended the methodology of Ng et al. (2006) by considering moving-average errors. Last, Same et al. (2011) suggested the use of mixture of linear experts for the clustering of electrical power consumption data.

This article is motivated by the problem of segmenting data that arise from the calcium imaging of zebrafish brains; see Muto and Kawakami (2013) for example. In such experiments, images containing tens-to-hundreds of thousands of pixels are obtained, where each of the individual pixels are time-series that may be thousands of periods long. Due to the size of the data, the joint modeling of both the spatial and temporal dependence of the data units is not possible. As such, we present a two-stage procedure: a temporal clustering stage (Stage 1), and a spatial smoothing stage (Stage 2).

In Stage 1, we perform temporal clustering via a mixture of autoregressions (MoAR) model. The report of our methodology is superficially similar to that of Xiong and Yeung (2004). We shall elaborate on key differences, both methodological and philosophical. First, the MoARMA model of Xiong and Yeung (2004) is fitted via maximum likelihood (ML) under the assumption that the data units are independent. Second, the EM (expectation–maximization) algorithm constructed only approximates the maximization of the moving-average (MA) parameters at each M-step; the ML estimation of MA models is well-known to be difficult (cf. Box et al. (2008, Sec. 7.3)). This implies that the algorithm does not have the usual numerical guarantees of an EM algorithm, such as monotonicity in likelihood evaluations and convergence to a stationary point of the log-likelihood function; see McLachlan and Krishnan (2008) regarding the properties of EM algorithms.

Unlike Xiong and Yeung (2004), we construct our MoAR model from the initial premise that the data units are spatially dependent, although we do not specify a joint distribution of the data units. Because of this, we cannot conduct ML estimation, since the likelihood function is unknown. We instead utilize a maximum marginal likelihood (MMaL) estimation approach, as described in Varin (2008). Conditions for the probabilistic consistency of the MMaL estimator are established.

Unfortunately, since the likelihood function is not used as the optimization objective, we cannot construct an EM algorithm due to the lack of a probabilistic model under which to compute conditional expectations. Fortunately, we can construct an algorithm via the MM (minorization–maximization) paradigm that yields an iterative scheme for the computation of the MMaL estimate; see Lange (2013, Ch. 8) for details regarding MM algorithms. Further, we prove that the MM algorithm monotonically increases the marginal likelihood (MaL) value at each iteration, and is convergent to a stationary point of the log-MaL function.

In Stage 2, upon performing model-based clustering on the data units via the fitted MoAR model, we then smooth the clustering outcomes via a Markov random field (MRF) model. The MRF model that we use can be viewed as a multivariate version of the one that is used in Nguyen et al. (2014). To fit the MRF model, we utilize a maximum pseudolikelihood (MPL) estimation approach; see Geman and Graffigne (1986) regarding the MPL estimation of MRF models. Again, an MM algorithm is constructed for the iterative computation of the MPL estimate. This MM algorithm is also proved to monotonically increase the pseudolikelihood (PL) value at each iteration, and is convergent to the global-maximum of the log-PL function. We also establish the consistency of the MPL estimator.

We note that the use of MRFs and mixture models for the joint modeling of spatial and temporal dependency in imaging data is not novel in its entirety; see for example Hartvig and Jensen (2000), Woolrich et al. (2005), and Vincent et al. (2010). Although interesting, these examples share some common shortcomings. First, in each of the examples, other than a probabilistic construction, no proofs are presented pertaining to the correctness of statistical inference that arise from the respective methodologies. Furthermore, each of the example models are either fitted using a Bayesian or an ad hoc estimation technique, that have unknown numerical properties. Last, the example methodologies are not suitable for the large-data nature of the zebrafish images; for example, Hartvig and Jensen (2000) only consider fMRI based time-series of lengths as short as 96 periods.

Besides the derivation of algorithms and proofs of theoretical results, we also provide a numerical simulation study that assesses the performance of our methodology. A calcium image of a zebrafish brain is then segmented to demonstrate the application of our methodology. The article proceeds as follows.

In Section 2, we introduce the MoAR model and derive an MM algorithm for the MMaL estimation of its model parameter. Here, the numerical properties of the MM algorithm are established, as well as the statistical properties of the MMaL estimator. In Section 3, we introduce the MRF model that is used for the smoothing of the clustering outcome. Here we also derive an MM algorithm for the MPL estimation of its model parameter. Further, the numerical properties of the MM algorithm are established, including the consistency of the MPL estimator. In Section 4, the numerical simulations are described, and results from the simulations are presented. In Section 5, we report on an example zebrafish brain calcium image segmentation. In Section 6, conclusions are drawn.

## 2 Mixture of autoregressions Models

Let $\boldsymbol{Y}_s = (Y_{s1}, ..., Y_{sm})^T \in \mathbb{R}^m$ be a random $m$-length time-series that is observed at the spatial position of the data unit $s = 1, ..., n$. Also let $Z_s \in \{1, ..., g\}$ be a latent random variable, where $\mathbb{P}(Z_s = i) = \pi_i > 0$ for $i = 1, ..., g$, and $\sum_{i=1}^{g} \pi_i = 1$. Let the probability density of $Y_{st} | \boldsymbol{Y}_{s(t)} = \boldsymbol{y}_{s(t)}, Z_s = i$ have the form

$$f\left(y_{st} | \boldsymbol{Y}_{s(t)} = \boldsymbol{y}_{s(t)}, Z_s = i\right) = \phi\left(y_{st}; \boldsymbol{y}_{s(t)}^T \boldsymbol{\beta}_i, \sigma_i^2\right), \tag{1}$$

where $t = p+1, p+2, ..., m$, $\boldsymbol{y}_{s(t)} = (1, y_{s,t-1}, ..., y_{s,t-p})^T$, $\boldsymbol{\beta}_i = (\beta_{i0}, ..., \beta_{ip})^T \in \mathbb{R}^{p+1}$, and $\sigma_i^2 > 0$ for each $i$. Then, we say that $\boldsymbol{Y}_s$ arises from a $g$-component MoAR model of order $p$ (or MoAR $(g, p)$, for brevity). Here, lower-case letters indicate realizations of random variables, upper-case letters indicate random variables, superscript $T$ denotes matrix transpositions, and

$$\phi\left(y; \mu, \sigma^2\right) = \left(2\pi\sigma^2\right)^{-1/2} \exp\left[-\frac{(x-\mu)^2}{2\sigma^2}\right]$$

is the normal density function with mean $\mu$ and variance $\sigma^2$.

Using characterization (1) and assuming that $Y_{s1}, ..., Y_{sp}$ are non-stochastic, we can deduce the conditional and marginal density characterizations of the MoAR $(g, p)$ model,

$$f\left(\boldsymbol{y}_s | Z_s = i; \boldsymbol{\theta}\right) = \prod_{t=p+1}^{m} \phi\left(y_{st}; \boldsymbol{y}_{s(t)}^T \boldsymbol{\beta}_i, \sigma_i^2\right)$$

and

$$f\left(\boldsymbol{y}_s; \boldsymbol{\theta}\right) = \sum_{i=1}^{g} \pi_i \prod_{t=p+1}^{m} \phi\left(y_{st}; \boldsymbol{y}_{s(t)}^T \boldsymbol{\beta}_i, \sigma_i^2\right), \qquad (2)$$

respectively. Here $\boldsymbol{\theta} = \left(\pi_1, ..., \pi_{g-1}, \boldsymbol{\beta}_1^T, ..., \boldsymbol{\beta}_g^T, \sigma_1^2, ..., \sigma_g^2\right)^T$ is the model parameter vector.

## 2.1 Maximum Marginal Likelihood Estimation

Suppose that $\boldsymbol{y}_1, ..., \boldsymbol{y}_n$ is a realization of the identically distributed (ID) sample of time-series $\boldsymbol{Y}_1, ..., \boldsymbol{Y}_n$, from a population with marginal densities (2). We assume that $\boldsymbol{Y}_1, ..., \boldsymbol{Y}_n$ are dependent, although we do not specify a joint distribution. As such, we cannot construct a likelihood function from the sample.

Following the approach of Varin (2008), we construct the MaL and log-MaL

functions, given by $\mathcal{M}_n(\boldsymbol{\theta}) = \prod_{s=1}^n f(\boldsymbol{y}_s; \boldsymbol{\theta})$ and

$$\begin{aligned}\ell_{\mathcal{M},n}(\boldsymbol{\theta}) &= \log \mathcal{M}_n(\boldsymbol{\theta}) \\ &= \sum_{s=1}^n \log \sum_{i=1}^g \pi_i \prod_{t=p+1}^m \phi\left(y_{st}; \boldsymbol{y}_{s(t)}^T \boldsymbol{\beta}_i, \sigma_i^2\right),\end{aligned} \quad (3)$$

respectively. The MMaL estimator can be defined as an appropriate local maximizer of (3) and is denoted by $\hat{\boldsymbol{\theta}}_n$.

Due to the log-summation form of (3), we cannot obtain $\hat{\boldsymbol{\theta}}_n$ as a root of the first-order condition $\nabla \ell_{\mathcal{M},n} = \mathbf{0}$ in closed form, where $\nabla$ is the gradient operator and $\mathbf{0}$ is a vector of zeros. As such, we require an iterative scheme for the computation of $\hat{\boldsymbol{\theta}}_n$.

## 2.2 Minorization–Maximization Algorithms

Suppose that $\ell(\boldsymbol{\theta})$ is an objective function that we wish to maximize, where $\boldsymbol{\theta} \in \Theta \subset \mathbb{R}^q$. If $\ell$ is difficult to maximize directly, then we can maximize a sequence of local approximations of $\ell$ instead. Let $\mathcal{U}(\boldsymbol{\theta}; \boldsymbol{\psi})$ be a minorizer of $\ell$ at $\boldsymbol{\psi} \in \Theta$ (we say that $\mathcal{U}$ minorizes $\ell$), where $\mathcal{U}$ is defined as follows.

**Definition 1.** $\mathcal{U}(\boldsymbol{\theta}; \boldsymbol{\psi})$ is a minorizer of $\ell(\boldsymbol{\theta})$, for $\boldsymbol{\theta}, \boldsymbol{\psi} \in \Theta \subset \mathbb{R}^q$, if $\ell(\boldsymbol{\theta}) = \mathcal{U}(\boldsymbol{\theta}; \boldsymbol{\theta})$ and $\ell(\boldsymbol{\theta}) \geq \mathcal{U}(\boldsymbol{\theta}; \boldsymbol{\psi})$.

Let $\boldsymbol{\theta}^{(0)}$ be some initialization and let $\boldsymbol{\theta}^{(r)}$ denote the $r$th iterate. An MM algorithm for the maximization of $\ell(\boldsymbol{\theta})$ via the minorizer $\mathcal{U}$ can be defined by the update rule

$$\boldsymbol{\theta}^{(r+1)} = \arg\max_{\boldsymbol{\theta} \in \Theta} \mathcal{U}\left(\boldsymbol{\theta}; \boldsymbol{\theta}^{(r)}\right). \quad (4)$$

Using Definition 1, we get the following result regarding (4).

**Proposition 1.** *If $\boldsymbol{\theta}^{(r)}$ is a sequence that is generated via the update rule (4), then the sequence $\ell\left(\boldsymbol{\theta}^{(r)}\right)$ is monotonically increasing with $r$.*

*Proof.* At each iteration $r$, Definition 1 and (4) imply the inequalities

$$\ell\left(\boldsymbol{\theta}^{(r+1)}\right) \geq \mathcal{U}\left(\boldsymbol{\theta}^{(r+1)}; \boldsymbol{\theta}^{(r)}\right) \geq \mathcal{U}\left(\boldsymbol{\theta}^{(r)}; \boldsymbol{\theta}^{(r)}\right) = \ell\left(\boldsymbol{\theta}^{(r)}\right).$$

$\square$

By Proposition 1, any MM algorithm will monotonically increase the objective function $\ell(\boldsymbol{\theta})$ at each iteration. The following minorizers from Lange (2013, Ch. 8) are useful in the construction of our algorithms.

**Fact 1.** *If $\Theta = [0, \infty)^q$, then $\ell(\boldsymbol{\theta}) = \log\left(\sum_{i=1}^{q} \theta_i\right)$ can be minorized by*

$$\mathcal{U}(\boldsymbol{\theta}; \boldsymbol{\psi}) = \sum_{i=1}^{q} \tau_i(\boldsymbol{\psi}) \log(\theta_i) - \sum_{i=1}^{q} \tau_i(\boldsymbol{\psi}) \log \tau_i(\boldsymbol{\psi}),$$

*where $\tau_i(\boldsymbol{\psi}) = \psi_i / \sum_{j=1}^{d} \psi_j$.*

**Fact 2.** *If $\Theta \subset \mathbb{R}^q$ and $h(\boldsymbol{\theta})$ is a function with Hessian $\mathcal{H}h(\boldsymbol{\theta})$ such that $\boldsymbol{H} - \mathcal{H}h(\boldsymbol{\theta})$ is negative semidefinite, for all $\boldsymbol{\theta}$, then $\ell(\boldsymbol{\theta}) = h(\boldsymbol{\theta})$ can be minorized by*

$$\mathcal{U}(\boldsymbol{\theta}; \boldsymbol{\psi}) = h(\boldsymbol{\psi}) + (\boldsymbol{\theta} - \boldsymbol{\psi})^T \nabla h(\boldsymbol{\psi}) + \frac{1}{2}(\boldsymbol{\theta} - \boldsymbol{\psi})^T \boldsymbol{H} (\boldsymbol{\theta} - \boldsymbol{\psi}).$$

### 2.3 MM Algorithm for MoAR Models

Starting from some initialization $\boldsymbol{\theta}^{(0)}$ and conditioning on the $r$th iterate $\boldsymbol{\theta}^{(r)}$, we get the following result via an application of Fact 1.

**Proposition 2.** *Given $\boldsymbol{\theta}^{(r)}$, the log-MaL function (3) can be minorized by*

$$\begin{aligned}
\mathcal{U}_{\mathcal{M}}\left(\boldsymbol{\theta};\boldsymbol{\theta}^{(r)}\right) &= \sum_{i=1}^{g}\sum_{s=1}^{n}\tau_{is}\left(\boldsymbol{\theta}^{(r)}\right)\left[\log\pi_i + \sum_{t=p+1}^{m}\log\phi\left(y_{st};\boldsymbol{y}_{s(t)}^T\boldsymbol{\beta}_i,\sigma_i^2\right)\right] \\
&\quad -\sum_{i=1}^{g}\sum_{s=1}^{n}\tau_{is}\left(\boldsymbol{\theta}^{(r)}\right)\log\tau_{is}\left(\boldsymbol{\theta}^{(r)}\right) \\
&= \sum_{i=1}^{g}\log\pi_i\sum_{s=1}^{n}\tau_{is}\left(\boldsymbol{\theta}^{(r)}\right) \qquad\qquad\qquad\qquad\qquad (5)\\
&\quad -\frac{1}{2}\sum_{i=1}^{g}\log\sigma_i^2\sum_{s=1}^{n}\sum_{t=p+1}^{m}\tau_{is}\left(\boldsymbol{\theta}^{(r)}\right) \\
&\quad -\frac{1}{2}\sum_{i=1}^{g}\frac{1}{\sigma_i^2}\sum_{s=1}^{n}\tau_{is}\left(\boldsymbol{\theta}^{(r)}\right)\sum_{t=p+1}^{m}\left(y_{st}-\boldsymbol{y}_{s(t)}^T\boldsymbol{\beta}_i\right)^2 + \mathcal{C}\left(\boldsymbol{\theta}^{(r)}\right),
\end{aligned}$$

*where $\tau_{is}\left(\boldsymbol{\theta}\right) = \pi_i f\left(\boldsymbol{y}_s|Z_s=i;\boldsymbol{\theta}\right)/f\left(\boldsymbol{y}_s;\boldsymbol{\theta}\right)$ and $\mathcal{C}\left(\boldsymbol{\theta}^{(r)}\right)$ collects up terms that do not depend on the active parameter $\boldsymbol{\theta}$.*

*Proof.* Make the substitution $\theta_i = \pi_i f\left(\boldsymbol{y}_s|Z_s=i;\boldsymbol{\theta}\right)$, for each $i$ and $s$, in Fact 1. □

To maximize (5) under the restriction $\sum_{i=1}^{g}\pi_i = 1$, we construct the Lagrangian $\Lambda\left(\boldsymbol{\theta},\lambda\right) = \mathcal{U}_{\mathcal{M}}\left(\boldsymbol{\theta};\boldsymbol{\theta}^{(r)}\right) + \lambda\left(\sum_{i=1}^{g}\pi_i - 1\right)$ and solve the first-order condition $\nabla\Lambda = \boldsymbol{0}$. This yields the updates

$$\pi_i^{(r+1)} = n^{-1}\sum_{s=1}^{n}\tau_{is}\left(\boldsymbol{\theta}^{(r)}\right), \qquad (6)$$

$$\boldsymbol{\beta}_i^{(r+1)} = \left[\sum_{s=1}^{n}\tau_{is}\left(\boldsymbol{\theta}^{(r)}\right)\sum_{t=p+1}^{m}\boldsymbol{y}_{s(t)}\boldsymbol{y}_{s(t)}^T\right]^{-1}\left[\sum_{s=1}^{n}\tau_{is}\left(\boldsymbol{\theta}^{(r)}\right)\sum_{t=p+1}^{m}\boldsymbol{y}_{s(t)}y_{st}\right], \qquad (7)$$

and

$$\sigma_i^{2(r+1)} = \frac{\sum_{s=1}^{n} \tau_{is}\left(\boldsymbol{\theta}^{(r)}\right) \sum_{t=p+1}^{m}\left(y_{st} - \boldsymbol{y}_{s(t)}^T \boldsymbol{\beta}_i^{(r+1)}\right)^2}{(m-p)\sum_{s=1}^{n}\tau_{is}\left(\boldsymbol{\theta}^{(r)}\right)}, \qquad (8)$$

for each $i$. Closely following the proof of Nguyen and McLachlan (2015, Thm. 2), we get the following result.

**Proposition 3.** *Given $\boldsymbol{\theta}^{(r)}$, if $\boldsymbol{\theta}^{(r+1)}$ is obtained via the updates (6)–(8) and*

$$\Theta = \left\{\pi_1, ..., \pi_{g-1} : \sum_{i=1}^{g-1} \pi_i < 1, \pi_i > 0 \right\} \times \mathbb{R}^{g(p+1)} \times (0, \infty)^g, \qquad (9)$$

*then*

$$\boldsymbol{\theta}^{(r+1)} = \arg\max_{\boldsymbol{\theta}\in\Theta} \mathcal{U}_\mathcal{M}\left(\boldsymbol{\theta}; \boldsymbol{\theta}^{(r)}\right).$$

Thus, Propositions 1–3 together imply that the MM algorithm defined via the updates (6)–(8) will monotonically increase the log-MaL at each iteration.

## 2.4 Convergence Analysis

Starting from some initialization $\boldsymbol{\theta}^{(0)}$, the MM algorithm is iterated (via updates (6)–(8)) until some convergence criterion is met, whereupon the final iterate is declared the MMaL estimate $\hat{\boldsymbol{\theta}}_n$. In this article, we choose to use the absolute convergence criterion $\ell_{\mathcal{M},n}\left(\boldsymbol{\theta}^{(r+1)}\right) - \ell_{\mathcal{M},n}\left(\boldsymbol{\theta}^{(r)}\right) < \delta$, for some small $\delta > 0$; see (Lange, 2013, Sec. 11.5) regarding the relative merits of different convergence criteria.

Let $\boldsymbol{\theta}^* = \lim_{r\to\infty} \boldsymbol{\theta}^{(r)}$ be a finite limit-point of the MM algorithm, starting from some initialization $\boldsymbol{\theta}^{(0)}$. The following result is adapted from Razaviyayn et al. (2013, Thm. 1).

**Lemma 1.** *If $\mathcal{U}$ is a minorizer of $\ell$ and $\boldsymbol{\theta}^{(r)}$ is a sequence of updates that is generated via rule (4), starting from some initialization $\boldsymbol{\theta}^{(0)}$, then the finite limit-point $\boldsymbol{\theta}^*$ is a stationary point of $\ell$.*

Lemma 1, and Propositions 2 and 3 together yield the following result.

**Theorem 1.** *If $\boldsymbol{\theta}^*$ is a finite limit-point of the sequence $\boldsymbol{\theta}^{(r)}$, obtained via updates (6)–(8) and starting from some initialization $\boldsymbol{\theta}^{(0)}$, then $\boldsymbol{\theta}^*$ is a saddle-point or local-maximum of (3).*

Theorem 1 is a useful result since it is known that likelihood-like objectives of mixture models tend to be highly multimodal and often unbounded. Because of this, it is often good practice to perform multiple randomized initializations of $\boldsymbol{\theta}^{(0)}$ in order to obtain a good local-maximum of the log-MaL function. For example, one can utilize the procedures from McLachlan and Peel (2000, Sec. 2.12.2).

## 2.5 Statistical Inference

We now seek asymptotic results regarding the MMaL estimator. Although no specific joint distribution of the data is specified, we do require some restrictions on the structure of dependency. As such, we must assume that the sequence $\boldsymbol{Y}_1, ..., \boldsymbol{Y}_n$ is either ergodic, or strongly mixing (i.e. $\alpha$-mixing); see (White, 2001, Sec. 3.3) for definitions. By applying Lemmas 3 and 4 (see Appendix I), we get the following result regarding the consistency of $\hat{\boldsymbol{\theta}}_n$.

**Theorem 2.** *Let $\boldsymbol{Y}_1, ..., \boldsymbol{Y}_n$ be an ID and ergodic (or $\alpha$-mixing) random sample, such that for each $s$, $\boldsymbol{Y}_s$ arises from a population with density function $f(\boldsymbol{y}_s; \boldsymbol{\theta}_0)$, where $\boldsymbol{\theta}_0$ is a strict-local maximizer of $\mathbb{E} \log f(\boldsymbol{Y}_s; \boldsymbol{\theta})$. If $\Theta_n = \{\boldsymbol{\theta} : \nabla \ell_{\mathcal{M}} = \mathbf{0}\}$ (where we take $\Theta_n = \{\bar{\boldsymbol{\theta}}\}$, for some $\bar{\boldsymbol{\theta}} \in \Theta$, if $\nabla \ell_{\mathcal{M}} = \mathbf{0}$ has no solution), then for any $\epsilon > 0$,*

$$\lim_{n \to \infty} \mathbb{P}\left[\inf_{\boldsymbol{\theta} \in \Theta_n} (\boldsymbol{\theta} - \boldsymbol{\theta}_0)^T (\boldsymbol{\theta} - \boldsymbol{\theta}_0) > \epsilon\right] = 0.$$

The proof of Theorem 2 appears in Appendix I. Theorem 2 is a useful result due to the lack of identifiability in mixture models (cf. Titterington et al. (1985, Ch. 3)). The theorem guarantees that at there exists a consistent sequence of strict-local maximizers of the log-MaL function.

We can obtain an asymptotic normality result, via Amemiya (1985, Thm. 4.1.3), although such a result is not useful in this article. The ergodicity (or mixing) assumption in Theorem 2 is quite general and offers little insight regarding the potential dependency structure in the data. The following result offers some intuition regarding the potential dependency structures in the data; see White (2001, Example 3.43) (cf. Ibragimov and Linnik (1971, Sec. 17.3)).

**Proposition 4.** *Let $\boldsymbol{Y}_t$ be a random sequence, such that $\boldsymbol{Y}_t$ is independent of $\boldsymbol{Y}_{t+\bar{t}}$ for each $\bar{t}$, where $0 < |\bar{t}| \leq \chi$. If $\chi < \infty$, then $\boldsymbol{Y}_t$ is strongly mixing.*

Proposition 4 implies that we need only assume that each data unit is limited to being dependent on a finite number of other data units, in order to fulfill the required mixing assumption. In practice, this result is sufficient justification for the consistency of $\hat{\boldsymbol{\theta}}_n$.

Following the approach of McLachlan and Basford (1988), we say that

$$\hat{c}_{sn} = \arg \max_{i=1,...,g} \tau_{is}\left(\hat{\boldsymbol{\theta}}_n\right) \qquad (10)$$

is the cluster allocation of data unit $s$. Via continuous mapping, if $\hat{\boldsymbol{\theta}}_n$ is a consistent estimator of $\boldsymbol{\theta}_0$, then $\hat{c}_{sn} \to c_s$ as $n \to \infty$, where $c_s$ is the Bayes' optimal allocation of data unit $s$ (cf. McLachlan (1992, Sec. 1.4)).

## 2.6 Model Selection

In all of our preceding discussions, it has been assumed that the number of components $g$ and the order $p$ have been fixed. If $g$ or $p$ are unknown in an MoAR $(g, p)$ model, it is not possible to determine their values via the previously presented MMaL estimation process.

In Xiong and Yeung (2004), an information-theoretic rule, based on the BIC (Bayesian information criterion; see Schwarz (1978)), was considered for the estimation of $g$ and $p$. Unfortunately, due to the lack of a likelihood function, this is not applicable in our case. Fortunately, we can utilize the PLIC (Pseudolikelihood information criterion; see Stanford and Raftery (2002)) as an approximate alternative. The PLIC rule can be described as follows.

Suppose that $g_0 \in \{\gamma_1, ..., \gamma_{m_g}\}$ is the true value of $g$, and $p_0 \in \{\psi_1, ..., \psi_{m_p}\}$ is the true value of $p$. For each pair $(k, l)$, where $k = 1, ..., m_g$ and $l = 1, ..., m_p$, we fit an MoAR $(\gamma_k, \psi_l)$ model via MMaL estimation to obtain the parameter estimates $\hat{\boldsymbol{\theta}}_{(k,l)n}$; the PLIC for the model can be computed as

$$\text{PLIC}_\mathcal{M}(k, l) = -2\ell_{\mathcal{M},n}\left(\hat{\boldsymbol{\theta}}_{(k,l)n}\right) + [g(p+3) - 1]\log n,$$

where $g(p+3) - 1$ is the number parameter components in $\hat{\boldsymbol{\theta}}_{(k,l)n}$. The PLIC rule for model selection is to set $g = \gamma_{\hat{k}}$ and $p = \psi_{\hat{l}}$, where

$$\left(\hat{k}, \hat{l}\right) = \arg\min_{k,l} \text{PLIC}_\mathcal{M}(k, l). \qquad (11)$$

## 3  Markov Random Field

Suppose that $c_1, ..., c_n$ is a realization of a sample of spatially dependent random variables $C_1, ..., C_n$ with unknown dependency structure. Let $\boldsymbol{w}_s$ be the spatial location of data unit $s$ (e.g. in a two-dimensional image, $\boldsymbol{w}_s = (w_s^x, w_s^y)^T$, $w_s^x$ is the $x$-coordinate and $w_s^y$ is the $y$-coordinate of the pixel), and let $\mathbb{C}_s^d = \{s' \neq s : \delta(\boldsymbol{w}_s, \boldsymbol{w}_{s'}) \leq d\}$ be a $d$-range neighborhood around unit $s$, where $\delta(\boldsymbol{w}_s, \boldsymbol{w}_{s'})$ is some distance between $\boldsymbol{w}_s$ and $\boldsymbol{w}_{s'}$. We take $\delta(\boldsymbol{w}_s, \boldsymbol{w}_{s'}) = \max\{|w_s^x - w_{s'}^x|, |w_s^y - w_{s'}^y|\}$ for two-dimensional images.

Let $C_{(s)} = \{C_{s'} : s' \in \mathbb{C}_s^d\}$ and $\eta_i\left(c_{(s)}\right) = \left|\mathbb{C}_s^d\right|^{-1} \sum_{s' \in \mathbb{C}_s^d} \mathbb{I}\left(c_{s'} = i\right)$; an approximation to the dependency structure of the sample can be made via the MRF characterization

$$\mathbb{P}\left(c_s = i | C_{(s)} = c_{(s)}; \boldsymbol{\psi}\right) = \prod_{i=1}^{g} \left[\frac{\exp\left(\boldsymbol{c}_{is}^T \boldsymbol{\psi}_i\right)}{\sum_{i'=1}^{g} \exp\left(\boldsymbol{c}_{i's}^T \boldsymbol{\psi}_{i'}\right)}\right]^{\mathbb{I}(c_s=i)}, \quad (12)$$

where $\boldsymbol{c}_{is} = \left(1, \eta_i\left(c_{(s)}\right)\right)^T$ and $\boldsymbol{\psi} = \left(\boldsymbol{\psi}_1^T, ..., \boldsymbol{\psi}_g^T\right)^T$ is the model parameter vector, with $\boldsymbol{\psi}_i = (\psi_{i1}, \psi_{i2})^T \in \mathbb{R}^2$ for each $i = 1, ..., g-1$, and $\boldsymbol{\psi}_g = \mathbf{0}$. Here, $\mathbb{I}(x = y)$ is an indicator variable that takes value 1 if $x = y$ and 0 otherwise. The MRF model (12) can be seen as a multinomial version of the binary MRF that is used in Nguyen et al. (2014).

## 3.1 Maximum Pseudolikelihood Estimation

Suppose that the sample $C_1, ..., C_n$ is best approximated via a model of form (12) with some parameter $\boldsymbol{\psi}_0$. To infer the value of $\boldsymbol{\psi}_0$, we follow Geman and Graffigne (1986) and construct the PL and log-PL functions $\mathcal{P}_n(\boldsymbol{\psi}) = \prod_{s=1}^{n} \prod_{i=1}^{g} \mathbb{P}\left(c_s = i | C_{(s)} = c_{(s)}; \boldsymbol{\psi}\right)$ and

$$\ell_{\mathcal{P},n}(\boldsymbol{\psi}) = \log \mathcal{P}_n(\boldsymbol{\psi}) \qquad (13)$$
$$\sum_{s=1}^{n} \sum_{i=1}^{g} \mathbb{I}(c_s = i) \boldsymbol{c}_{is}^T \boldsymbol{\psi}_i - \sum_{s=1}^{n} \log \sum_{i'=1}^{g} \exp\left(\boldsymbol{c}_{i's}^T \boldsymbol{\psi}_{i'}\right),$$

respectively. We then estimate $\boldsymbol{\psi}_0$ via the MPL estimator

$$\hat{\boldsymbol{\psi}}_n = \arg\max_{\boldsymbol{\psi} \in \Psi} \ell_{\mathcal{P},n}(\boldsymbol{\psi}), \qquad (14)$$

where $\Psi = \mathbb{R}^{2(g-1)}$. We note that the first-order condition $\nabla \ell_{\mathcal{P},n} = \mathbf{0}$ does not have a closed-form expression. As such, we require an iterative method for the computation of (14).

## 3.2 Block-wise MM Algorithms

Using the same notation as Section 2.2, suppose that $\boldsymbol{\theta}$ can be partitioned into $k$ blocks, where $\boldsymbol{\theta} = \left(\boldsymbol{\theta}_1^T, ..., \boldsymbol{\theta}_k^T\right)^T$, where $\boldsymbol{\theta}_i \in \Theta_i$ for each $i = 1, ..., k$, and $\Theta = \prod_{i=1}^{k} \Theta_i$. We say that $\mathcal{U}_i\left(\boldsymbol{\theta}_i; \boldsymbol{\psi}\right)$ is the $i$th block-wise minorizer of $\ell$ if it fulfills the following definition.

**Definition 2.** $\mathcal{U}_i\left(\boldsymbol{\theta}_i; \boldsymbol{\psi}\right)$ is the $i$th block-wise minorizer of $\ell\left(\boldsymbol{\theta}\right)$, for $\boldsymbol{\theta}, \boldsymbol{\psi} \in \Theta$ and $\boldsymbol{\theta}_i \in \Theta_i$, if $\ell\left(\boldsymbol{\theta}\right) = \mathcal{U}_i\left(\boldsymbol{\theta}_i; \boldsymbol{\theta}\right)$ and $\ell\left(\boldsymbol{\theta}\right) \geq \mathcal{U}\left(\boldsymbol{\theta}_i; \boldsymbol{\psi}\right)$.

Let $\boldsymbol{\theta}^{(0)}$ be some initialization, and let $\boldsymbol{\theta}^{(r)}$ denote the $r$th iterate. A block-wise MM (BMM) algorithm for the maximization of $\ell$ via the block-wise minorizer $\mathcal{U}_1, ..., \mathcal{U}_k$ can be defined by the update rule

$$\boldsymbol{\theta}_i^{(r+1)} = \begin{cases} \arg\max_{\boldsymbol{\theta}_i \in \Theta_i} \mathcal{U}_i\left(\boldsymbol{\theta}_i; \boldsymbol{\theta}^{(r)}\right) & \text{if } i = (r \bmod k) + 1, \\ \boldsymbol{\theta}_i^{(r)} & \text{otherwise}, \end{cases} \quad (15)$$

for each $i$. Using Definition 1, we get the following result regarding (15).

**Proposition 5.** *If $\boldsymbol{\theta}^{(r)}$ is a sequence that is generated via the update rule (15), then the sequence $\ell\left(\boldsymbol{\theta}^{(r)}\right)$ is monotonically increasing in $r$.*

*Proof.* Without loss of generality, suppose that at iteration $r$, $i = (r \bmod k) + 1$. Then, Definition (2) and rule (15) imply

$$\ell\left(\boldsymbol{\theta}^{(r+1)}\right) \geq \mathcal{U}_i\left(\boldsymbol{\theta}^{(r+1)}; \boldsymbol{\theta}^{(r)}\right) \geq \mathcal{U}_i\left(\boldsymbol{\theta}^{(r)}; \boldsymbol{\theta}^{(r)}\right) = \ell\left(\boldsymbol{\theta}^{(r)}\right).$$

□

Thus, like MM algorithms, BMM algorithms also monotonically increase the value of the objective $\ell$ at each iteration.

### 3.3 BMM Algorithm for MRF Model

Starting from some initialization $\boldsymbol{\psi}^{(0)}$ and conditioned on the $r$th iterate $\boldsymbol{\psi}^{(r)}$, we get the following result via an application of Fact 2.

**Proposition 6.** *Given $\boldsymbol{\psi}^{(r)}$, for each $i = 1, ..., g-1$, the log-MaL function (13) can be block-wise minorized by*

$$\begin{aligned}\mathcal{U}_{\mathcal{P},i}\left(\boldsymbol{\psi}_i; \boldsymbol{\psi}^{(r)}\right) &= \ell_{\mathcal{P},n}\left(\boldsymbol{\psi}^{(r)}\right) + \left(\boldsymbol{\psi}_i - \boldsymbol{\psi}_i^{(r)}\right)^T \nabla_i \ell_{\mathcal{P},n}\left(\boldsymbol{\psi}^{(r)}\right) \\ &\quad - \frac{1}{8}\left(\boldsymbol{\psi}_i - \boldsymbol{\psi}_i^{(k)}\right)^T \boldsymbol{\Delta}_i \left(\boldsymbol{\psi}_i - \boldsymbol{\psi}_i^{(k)}\right),\end{aligned} \qquad (16)$$

*where*

$$\nabla_i \ell_{\mathcal{P},n}(\boldsymbol{\psi}) = \sum_{s=1}^n \mathbb{I}(v_s = i) \boldsymbol{c}_{is} - \sum_{s=1}^n \boldsymbol{c}_{is} \frac{\exp\left(\boldsymbol{c}_{is}^T \boldsymbol{\psi}_i\right)}{\sum_{i'=1}^g \exp\left(\boldsymbol{c}_{i's}^T \boldsymbol{\psi}_{i'}\right)}$$

*is the partial derivative of $\ell_{\mathcal{P},n}$ with respect to $\boldsymbol{\psi}_i$, and $\boldsymbol{\Delta}_i = \sum_{s=1}^n \boldsymbol{c}_{is} \boldsymbol{c}_{is}^T$.*

*Proof.* For each $i = 1, ..., g-1$, note that the second derivative with respect to $\boldsymbol{\psi}_i$, can be written as

$$\mathcal{H}_i \ell_{\mathcal{P},n} = -\sum_{s=1}^n \boldsymbol{c}_{is} \boldsymbol{c}_{is}^T p_{is}(1 - p_{is})$$

where $p_{is} = \exp\left(\boldsymbol{c}_{is}^T \boldsymbol{\psi}_i\right) / \sum_{i'=1}^g \exp\left(\boldsymbol{c}_{i's}^T \boldsymbol{\psi}_{i'}\right)$. Notice that $p_{is} \in (0,1)$, and thus $p_{is}(1 - p_{is}) \le 1/4$ (i.e. when $p_{is} = 1/2$). Thus, $-\boldsymbol{\Delta}_i/4 - \mathcal{H}_i \ell_{\mathcal{P},n}$ is negatively semidefinite. We therefore set $\boldsymbol{\theta} = \boldsymbol{\psi}_i$ and $\boldsymbol{H} = -\boldsymbol{\Delta}_i/4$ in Fact (2), for each $i$. □

By solving the first-order condition $\nabla \mathcal{U}_{\mathcal{P},i} = \boldsymbol{0}$ for each $i = 1, ..., g-1$, we get the update rule

$$\boldsymbol{\psi}_i^{(r+1)} = \begin{cases} 4\boldsymbol{\Delta}_i^{-1}\boldsymbol{p}_i\left(\boldsymbol{\psi}^{(k)}\right) + \boldsymbol{\psi}_i^{(r)} & \text{if } i = (r \bmod g - 1) + 1, \\ \boldsymbol{\psi}_i^{(r)} & \text{otherwise.} \end{cases} \quad (17)$$

Note that (16) is quadratic and thus $\boldsymbol{\psi}_i^{(r+1)}$ is the global maximum of (16), when $i = (r \bmod g - 1) + 1$. Thus Propositions 5 and 6 imply that the BMM algorithm defined via Rule (17) will monotonically increase the log-PL value at each iteration.

### 3.4 Convergence Analysis

Starting from some initialization $\boldsymbol{\psi}^{(0)}$, the BMM algorithm is iterated via update rule (17) until $\ell_{\mathcal{P},n}\left(\boldsymbol{\psi}^{(r+1)}\right) - \ell_{\mathcal{P},n}\left(\boldsymbol{\psi}^{(r)}\right) < \delta$, whereupon the final iterate is declared the MPL estimate $\hat{\boldsymbol{\psi}}_n$. Let $\boldsymbol{\psi}^* = \lim_{r\to\infty} \boldsymbol{\psi}^{(r)}$ be a finite limit-point of the BMM algorithm. The following result is adapted from Razaviyayn et al. (2013, Thm. 2).

**Lemma 2.** *For each $i = 1, ..., k$, if $\mathcal{U}_i$ is a block-wise minorizer of $\ell$ and $\boldsymbol{\theta}^{(r)}$ is a sequence of updates that is generated via rule (15), starting from some initialization $\boldsymbol{\theta}^{(0)}$, then the finite limit-point $\boldsymbol{\theta}^*$ is a stationary point of $\ell$.*

Lemma 2, and Proposition 6 together yield the following result.

**Theorem 3.** *If $\boldsymbol{\psi}^*$ is a finite limit-point of the sequence $\boldsymbol{\psi}^{(r)}$, obtained via updates (17) and starting from some initialization $\boldsymbol{\psi}^{(0)}$, then $\boldsymbol{\psi}^*$ is a stationary point of (13).*

Note that (13) is only nonlinear in the terms $\sum_{s=1}^{n} \log \sum_{i'=1}^{g} \exp\left(\boldsymbol{v}_{i's}^T \boldsymbol{\psi}_{i'}\right)$, which are convex since they are of log-sum-exp form (cf. Boyd and Vandenberghe (2004, Sec. 3.1)). Thus (13) is concave, and we have the following strengthening of Theorem (3).

**Corollary 1.** *The finite limit-point $\boldsymbol{\psi}^*$ from Theorem 3 is the global-maximizer of (13).*

## 3.5 Statistical Inference

We now seek asymptotic results regarding the MPL estimator. To attain such a result, we define the notion of identifiability.

**Definition 3.** Let $\boldsymbol{\psi}, \boldsymbol{\psi}' \in \mathbb{R}^{2g}$. If there exists $c_s$ and $c_{(s)}$, such that

$$\mathbb{P}\left(c_s = i | C_{(s)} = c_{(s)}; \boldsymbol{\psi}\right) \neq \mathbb{P}\left(c_s = i | C_{(s)} = c_{(s)}; \boldsymbol{\psi}'\right),$$

for all $\boldsymbol{\psi} \neq \boldsymbol{\psi}'$, then we say that the MRF (12) is identifiable.

In Appendix II we prove the identifiability of (12) and apply an adaptation of the Theorem from Geman and Graffigne (1986), to attain the following result.

**Theorem 4.** *If $C_1, ..., C_n$ is a random sample that is best approximated via a model of form (12) with some parameter $\boldsymbol{\psi}_0$, then $\hat{\boldsymbol{\psi}}_n \xrightarrow{P} \boldsymbol{\psi}_0$.*

By Theorem 4, (14) is a consistent estimator of the parameter $\boldsymbol{\psi}_0$. Let $\tilde{c}_{sn}$ be the smoothed cluster allocation of data unit $s$, where

$$\tilde{c}_{sn} = \arg\max_{i=1,...,g} \mathbb{P}\left(c_s = i | C_{(s)} = c_{(s)}; \hat{\boldsymbol{\psi}}_n\right). \tag{18}$$

Via continuous mapping, $\tilde{c}_{sn}$ is consistent with respect to the best approximate MRF allocation of the random sample $C_1, ..., C_n$.

## 3.6 Range Estimation

Thus far, the range $d$ of the neighborhood $\mathbb{C}_s^d$ has been assumed constant. It is intractable to estimate $d$ via the BMM algorithm for MPL estimation, described in Section 3.3. Thus, we require an auxiliary method for estimating $d$. As in Section 2.6, we utilize a PLIC-based method.

Suppose that $d_0 \in \{\delta_1, ..., \delta_{m_d}\}$ is the true value of $d$. For each $k = 1, ..., m_d$, we fit an MRF model via MPL estimation to obtain the parameter estimates $\hat{\boldsymbol{\psi}}_{(k)n}$; the PLIC for the model can be computed as

$$\text{PLIC}_{\mathcal{P}}(k) = -2\ell_{\mathcal{P},n}\left(\hat{\boldsymbol{\psi}}_{(k)n}\right) + (2g - 2)\log n,$$

where $2g - 2$ is the number parameter components in $\hat{\boldsymbol{\psi}}_{(k)n}$. The PLIC rule for model selection is to set $d = \delta_{\hat{k}}$, where

$$\hat{k} = \arg\min_{k} \text{PLIC}_{\mathcal{P}}(k). \tag{19}$$

Rule (19) is known to be consistent for choosing between competing MRF models (cf. (Ji and Seymour, 1996)).

## 4 Numerical Simulations

In order to assess the performance of our algorithms for the clustering of spatially dependent time-series, we perform a set of four different simulations S1–S4; we shall refer to Figure 1 in the descriptions in sequel. In all four scenarios, an $n = 100 \times 100$ image is simulated, where each pixel is a realization of an $m = 100$ long AR (autoregressive) time-series. In S1 and S2, dark blue and red pixels are realizations of classes C1 and C2 AR time-series, respectively. In S3 and S4, dark blue, red, light blue, and yellow pixels are realizations of classes C1–C4 AR time-series, respectively. We refer to Table 1 for the parameter vector of each class, and we graph three typical realizations of each class in Figure 2. We can interpret S1 and S2 as having arisen from MoAR $(2, 2)$ models, and we can interpret S3 and S4 as having arisen from $\mathrm{MoAR}\left(4,2\right)$ models.

For each scenario, we estimate $g=1,...,5$ components MoAR models of orders $p=1,...,5$ and record their PLIC values. We repeat this $N=100$ times for each case, and record the number of times each combination $(g,p)$ has the smallest PLIC value, as well as the average PLIC value for each combination. The results are recorded in Table 2–5.

Using the most often selected model, from the previous simulations, we compute the adjusted Rand index (ARI) of Hubert and Arabie (1985) to determine the similarity between the Stage 1 clustering of the pixels using Rule (10) and the true classes (see Figure 1). We then compare this with the ARI computed from the Stage 2 MRF clustering (i.e. via rule (18)) and the true classes. The ARI is a measure of concordance between two clusterings, where a value of 1 indicates perfect concordance, 0 indicates no relationship, and -1 indicates perfect discordance. We repeat the comparisons $N=100$ times and report the average ARI values of the two rules and the average neighborhood distance $d$ in Table 6; here, we select greedily select $d$ (cf. Nguyen et al. (2014)). Examples of corresponding Stage 1 and Stage 2 clusterings are plotted in Figures 3 and 4, respectively.

All of our simulations are conducted in the $R$ statistical programming environment (R Core Team, 2013). AR time-series are generated using the *arima.sim* function in $R$. The MM algorithms are programmed in $R$, with the log-MaL and log-PL values and MM algorithm updates coded in $C$ via the *Rcpp* and *RcppArmadillo* packages (Eddelbuettel, 2013). The ARI values are computed using the *adjustedRandIndex* function from the *mclust* $R$ package (Fraley and Raftery, 2002, 2003).

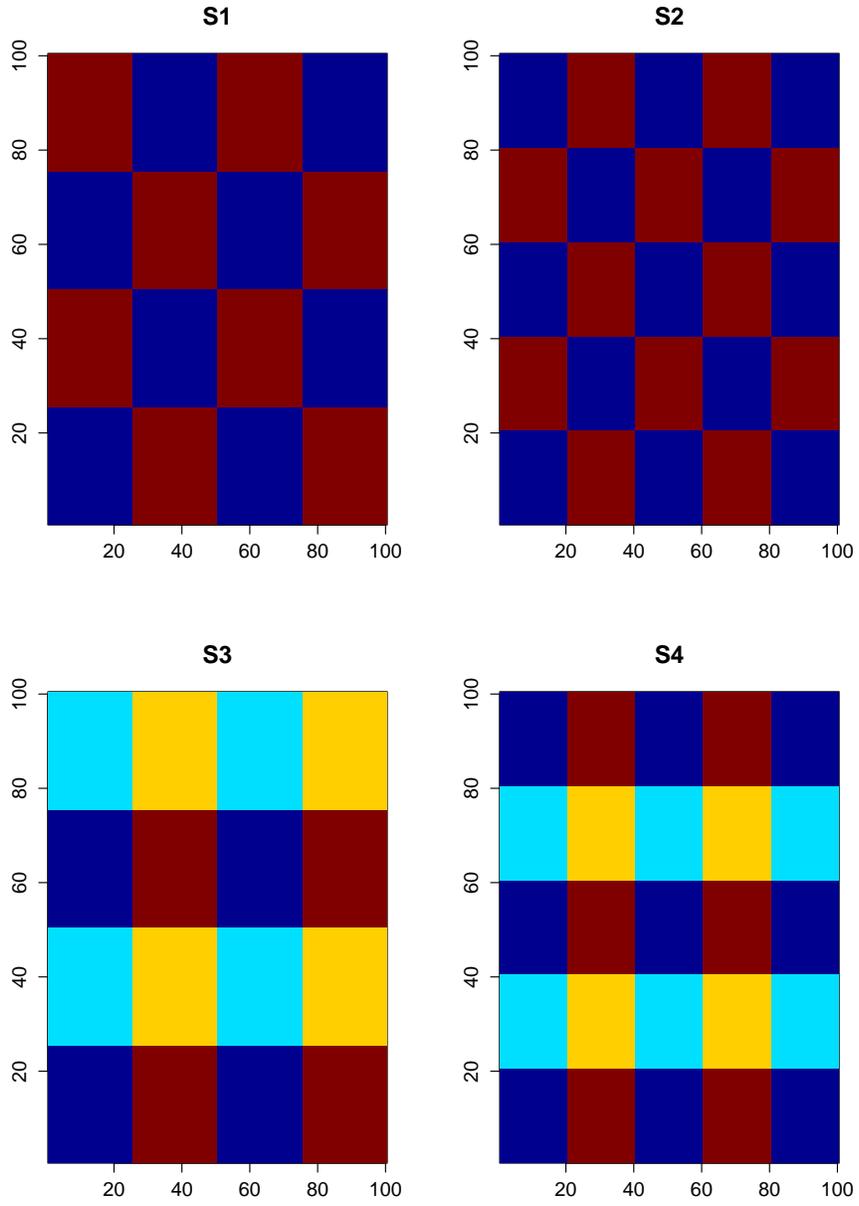

Figure 1: True classes for S1–S4. Pixels colored dark blue, red, light blue, and yellow are generated from classes C1–C4, respectively.

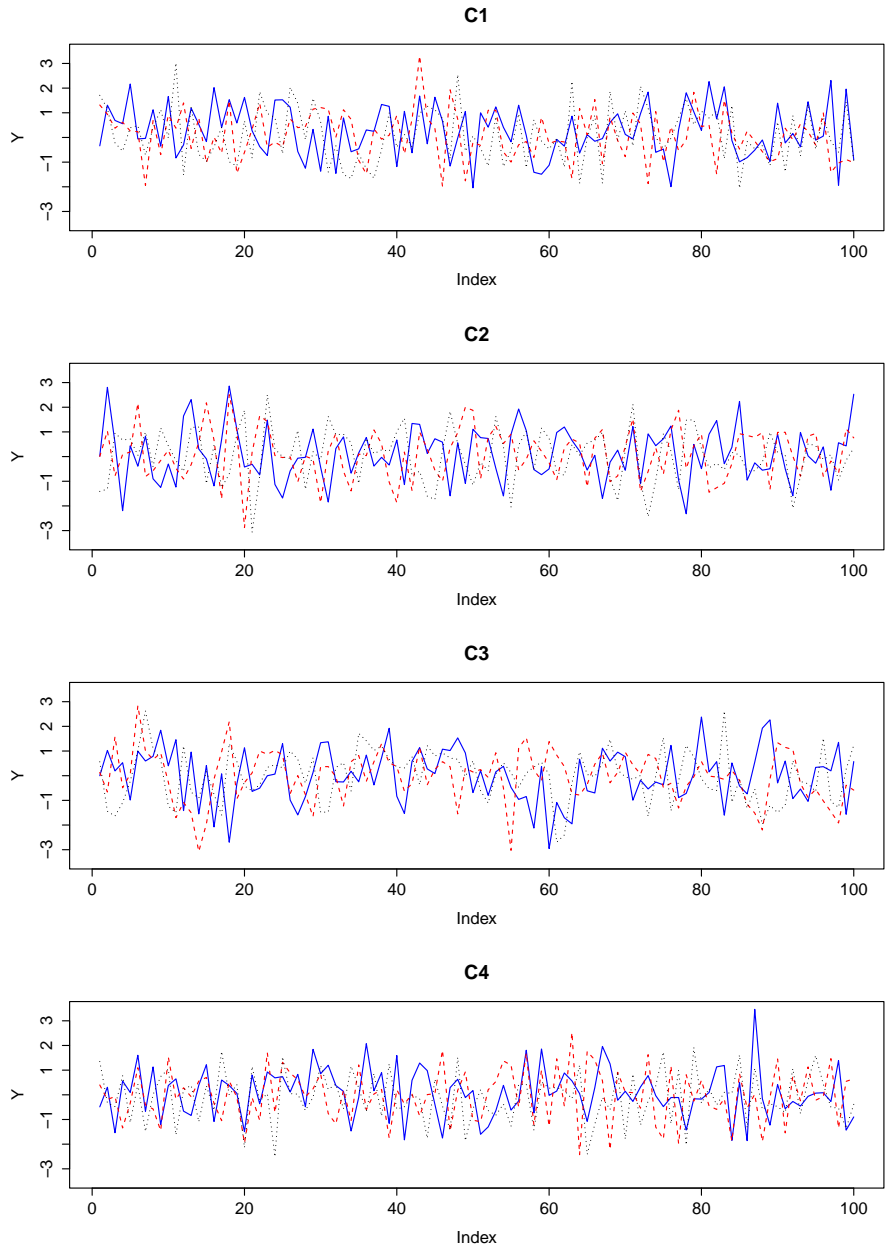

Figure 2: Typical realizations of time-series from classes C1–C4; see Table 1 regarding the generative model of each class.

Table 1: Parameter vectors of C1–C4, as used in S1–S4.

| Case (i.e $i$) | $\beta_{i0}$ | $\beta_{i1}$ | $\beta_{i2}$ | $\sigma_i^2$ |
|---|---|---|---|---|
| C1 ($i=1$) | 0 | 0 | 0.25 | 1 |
| C2 ($i=2$) | 0 | 0 | -0.25 | 1 |
| C3 ($i=3$) | 0 | 0.25 | 0 | 1 |
| C4 ($i=4$) | 0 | -0.25 | 0 | 1 |

Table 2: S1 Results. The "Average PLIC" column indicates the average value over $N = 100$ repetitions. The "Number Picked" column indicates the number of times Rule (11) selected the model. Underline indicates the correct model, bold indicates the most selected model.

| | Average PLIC | | | | | Number Picked | | | | |
|---|---|---|---|---|---|---|---|---|---|---|
| $g\|p$ | 1 | 2 | 3 | 4 | 5 | 1 | 2 | 3 | 4 | 5 |
| 1 | 2757365 | 2757373 | 2757381 | 2753655 | 2753663 | 0 | 0 | 0 | 0 | 0 |
| 2 | 2757402 | **<u>2709432</u>** | 2709449 | 2709466 | 2709482 | 0 | **<u>100</u>** | 0 | 0 | 0 |
| 3 | 2757351 | 2709498 | 2709524 | 2709550 | 2709576 | 0 | 0 | 0 | 0 | 0 |
| 4 | 2757361 | 2709566 | 2709603 | 2709640 | 2709676 | 0 | 0 | 0 | 0 | 0 |
| 5 | 2757387 | 2709654 | 2709701 | 2709750 | 2709797 | 0 | 0 | 0 | 0 | 0 |

Table 3: S2 Results. The "Average PLIC" column indicates the average value over $N = 100$ repetitions. The "Number Picked" column indicates the number of times Rule (11) selected the model. Underline indicates the correct model, bold indicates the most selected model.

| | Average PLIC | | | | | Number Picked | | | | |
|---|---|---|---|---|---|---|---|---|---|---|
| $g\|p$ | 1 | 2 | 3 | 4 | 5 | 1 | 2 | 3 | 4 | 5 |
| 1 | 2757385 | 2757295 | 2757303 | 2753585 | 2753593 | 0 | 0 | 0 | 0 | 0 |
| 2 | 2757422 | **<u>2709482</u>** | 2709499 | 2709515 | 2709531 | 0 | **<u>100</u>** | 0 | 0 | 0 |
| 3 | 2757348 | 2709546 | 2709573 | 2709599 | 2709625 | 0 | 0 | 0 | 0 | 0 |
| 4 | 2757348 | 2709613 | 2709650 | 2709687 | 2709723 | 0 | 0 | 0 | 0 | 0 |
| 5 | 2757373 | 2709705 | 2709752 | 2709801 | 2709849 | 0 | 0 | 0 | 0 | 0 |

Table 4: S3 Results. The "Average PLIC" column indicates the average value over $N = 100$ repetitions. The "Number Picked" column indicates the number of times Rule (11) selected the model. Underline indicates the correct model, bold indicates the most selected model.

| | Average PLIC | | | | | Number Picked | | | | |
|---|---|---|---|---|---|---|---|---|---|---|
| $g\|p$ | 1 | 2 | 3 | 4 | 5 | 1 | 2 | 3 | 4 | 5 |
| 1 | 2757133 | 2756201 | 2756210 | 2755224 | 2755232 | 0 | 0 | 0 | 0 | 0 |
| 2 | 2757170 | 2738857 | 2738868 | 2738867 | 2738885 | 0 | 0 | 0 | 0 | 0 |
| 3 | 2742374 | 2725611 | 2725356 | 2725312 | 2725338 | 0 | 0 | 0 | 0 | 0 |
| 4 | 2739632 | **<u>2719116</u>** | 2719149 | 2719181 | 2719214 | 0 | **<u>100</u>** | 0 | 0 | 0 |
| 5 | 2739672 | 2719243 | 2719288 | 2719332 | 2719377 | 0 | 0 | 0 | 0 | 0 |

Table 5: S4 Results. The "Average PLIC" column indicates the average value over $N = 100$ repetitions. The "Number Picked" column indicates the number of times Rule (11) selected the model. Underline indicates the correct model, bold indicates the most selected model.

| | Average PLIC | | | | | Number Picked | | | | |
|---|---|---|---|---|---|---|---|---|---|---|
| $g\|p$ | 1 | 2 | 3 | 4 | 5 | 1 | 2 | 3 | 4 | 5 |
| 1 | 2756774 | 2753931 | 2753939 | 2752697 | 2752704 | 0 | 0 | 0 | 0 | 0 |
| 2 | 2756811 | 2733417 | 2733407 | 2733401 | 2733418 | 0 | 0 | 0 | 0 | 0 |
| 3 | 2746471 | 2725642 | 2725350 | 2725329 | 2725353 | 0 | 0 | 0 | 0 | 0 |
| 4 | 2743711 | **2718329** | 2718362 | 2718394 | 2718427 | 0 | **100** | 0 | 0 | 0 |
| 5 | 2743748 | 2718448 | 2718493 | 2718537 | 2718581 | 0 | 0 | 0 | 0 | 0 |

Table 6: ARI values for clustering outcomes from the Stage 1 Rule (10) and the Stage 2 Rule (18), averaged over $N = 100$ repetitions. The Stage 1 column displays the results obtained using an MoAR $(2, 2)$ in S1 and S2, and MoAR $(4, 2)$ in S3 and S4. The Stage 2 column displays the results obtained upon MRF smoothing. The $d$ column reports the average size of the neighborhoods used in the MRF models.

| Scenario | Stage 1 | Stage 2 | $d$ |
|---|---|---|---|
| S1 | 0.970 | 0.998 | 1.02 |
| S2 | 0.970 | 0.997 | 1.02 |
| S3 | 0.759 | 0.947 | 1.00 |
| S4 | 0.723 | 0.917 | 1.00 |

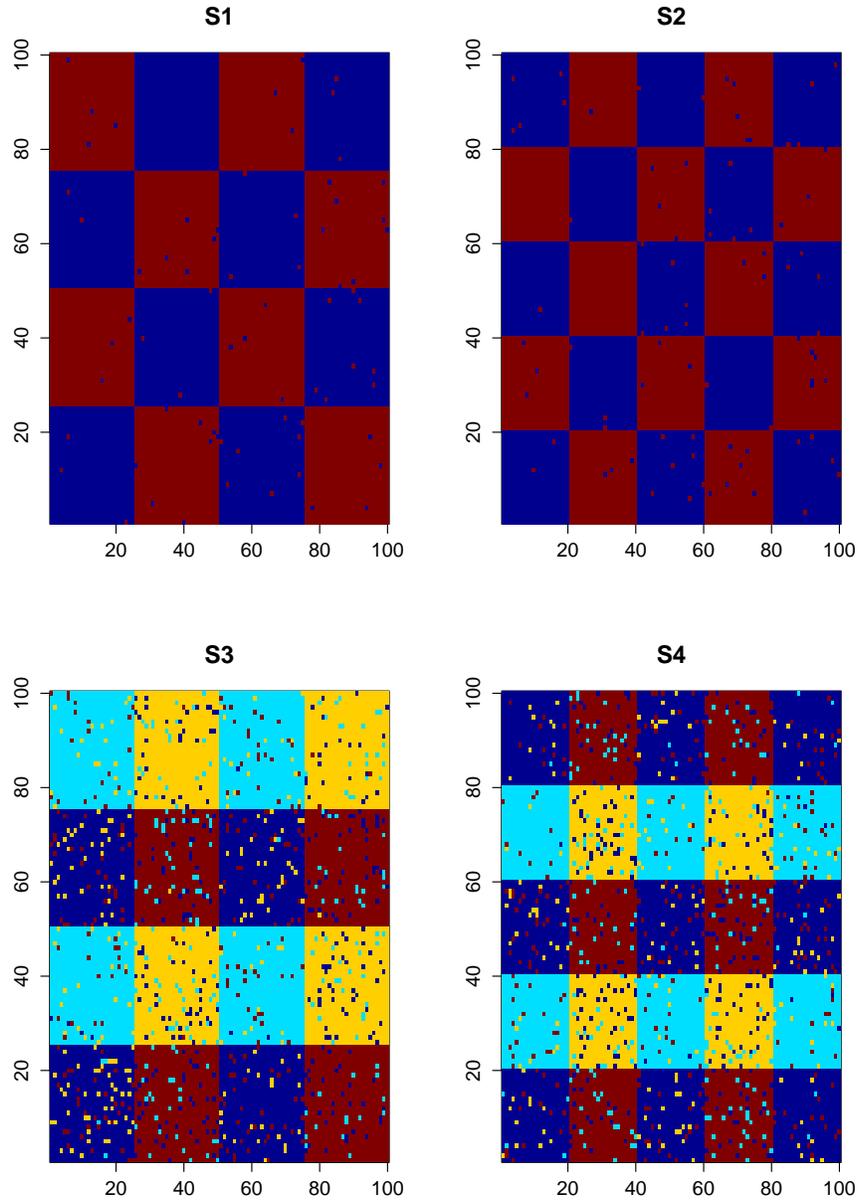

Figure 3: Example Stage 1 clusterings using an MoAR $(2, 5)$ in S1 and S2, and MoAR $(4, 5)$ in S3 and S4.

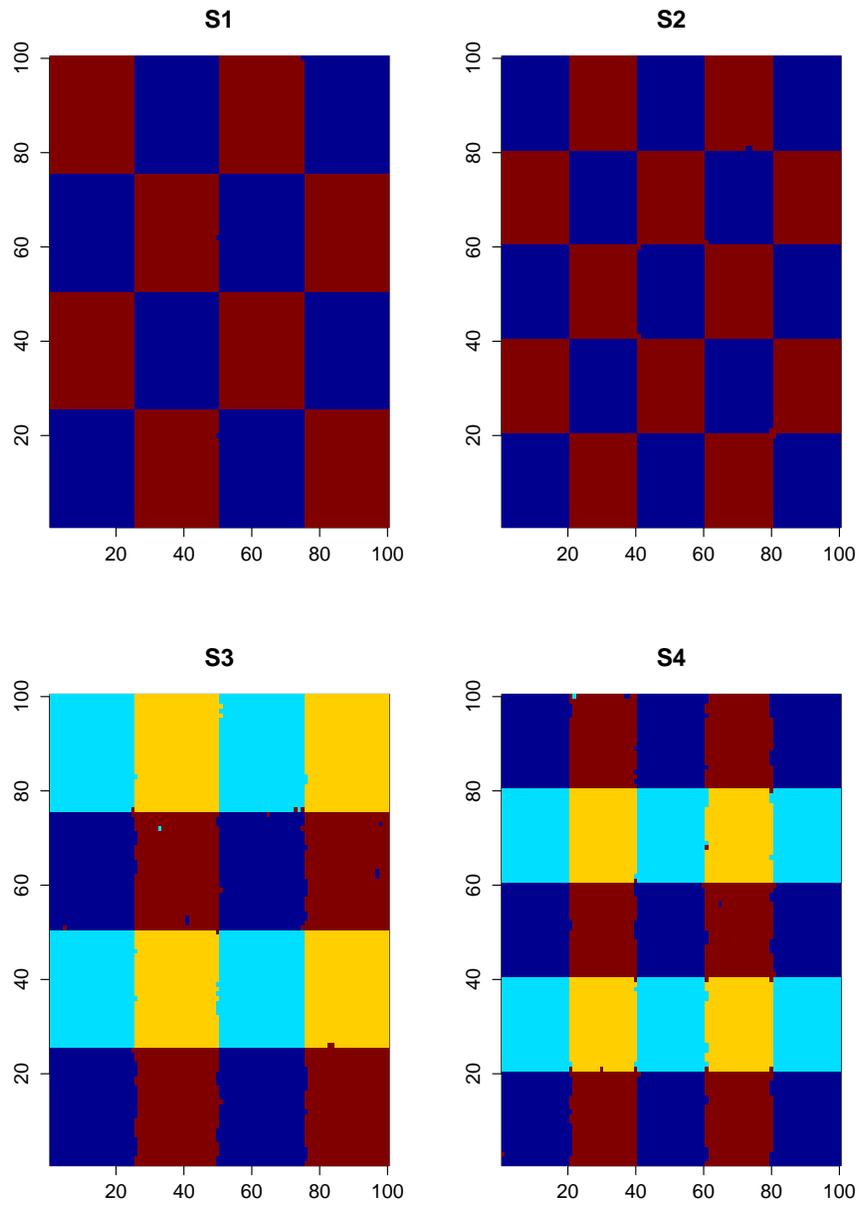

Figure 4: Example Stage 2 clusterings using the same models as in Figure 3, via MRFs with $d = 1$.

### 4.1 Discussion

We note that the classes C1–C4 were selected for two reasons. First, each AR model has mean zero, thus methods that could be implemented on the time-averaged data would not have been effective. This rules out methods such as $k$-means (MacQueen, 1967) or the normal mixture class of model-based clustering algorithms (e.g. *EMMIX (McLachlan et al., 1999)*; see also McLachlan and Basford (1988) and McLachlan and Peel (2000, Ch. 3)). Second, we selected the classes to be stationary AR models, implying that the means and variances stay constant over time (cf. Box et al. (2008, Ch. 2)). Thus, methods that estimate a mean functions would not be effective (e.g mixtures of regressions (DeSarbo and Cron, 1988; Jones and McLachlan, 1992) or mixtures of experts (Same et al., 2011)).

From Tables 2–5, we observe that the PLIC rule was able to select the correct number of clusters and correct order on every occasion. This affirms the appropriateness of the PLIC rule for this example.

The Stage 2 columns of Table 6 indicate that the use of an MRF model to account for spatial dependencies does drastically improve the performance of the clustering. All of the ARI values, after Stage 2, are close to one; this indicates a very strong concordance between the true classes and the clustering outcomes. Furthermore, the $d$ columns of the same table indicate that only small neighborhoods of dependencies are necessary in order to attain such improvements. Figures 3 and 4 further illustrate that the use of neighborhood dependencies result in significant improvements in the clustering outcomes.

## 5 Example Application

To demonstrate the use of the two-stage procedure, we consider an analysis of a time-series data set arising from the calcium imaging of a zebrafish brain. The calcium imaging was performed on a 5 day post fertilization GCaMP5 (Akerboom et al., 2012) zebrafish brain; see Muto and Kawakami (2013) for example, regarding the calcium imaging of zebrafish brains. The images were acquired on an inverted spinning disk microscope, more specifically, a Zeiss Axio Observer Z1 with a W1 Yokogawa spinning disk module and a Hamamatsu Flash 4.0 sCMOS camera. The zebrafish was subjected to pharmacologically induced neuronal activation over 500 seconds, and a single plane of $1024 \times 1024$ time-series data was acquired at 10 Hz over this time period. The data are down-sampled to a $512 \times 512$ image in space and the time series are smoothed and down-sampled to length $m = 500$ in time. Using a numerical threshold, the image is manually masked to produce a final set of $n = 23445$ spatially correlated time-series from pixels displaying interesting neuronal activity. A time-averaged image of the data is presented in Figure 5.

To segment the image, we estimate $g = 1, ..., 25$ components MoAR models of orders $p = 1, ..., 20$, where $g+p \leq 26$. A heatmap of the obtained PLIC values for the models that are considered is presented in Figure 6. A Stage 1 clustering using the best model from Figure 6 (i.e. MoAR $(17, 3)$ and $\text{PLIC}_\mathcal{M} = 2292875$) is presented in Figure 7, and a Stage 2 clustering, using an MRF with $d = 1$ ($\text{PLIC}_\mathcal{P}(1) = 42168.78$, $\text{PLIC}_\mathcal{P}(2) = 62110.58$), is presented in Figure 8. The Stage 1 clustering yielded 917, 923, 644, 632, 721, 847, 763, 765, 757, 756, 1037, 1551, 2345, 2979, 2672, 2788, and 2348 pixels in clusters 1–17, respectively. The Stage 2 clustering yielded 927, 922, 544, 592, 794, 630, 678, 632, 862, 873, 1031, 1476, 2325, 3097, 2664, 2921, and 2477 pixels in clusters 1–17, respectively. Altogether, the Stage 1 and Stage 2 clusterings matched on 71.8% of the pixels.

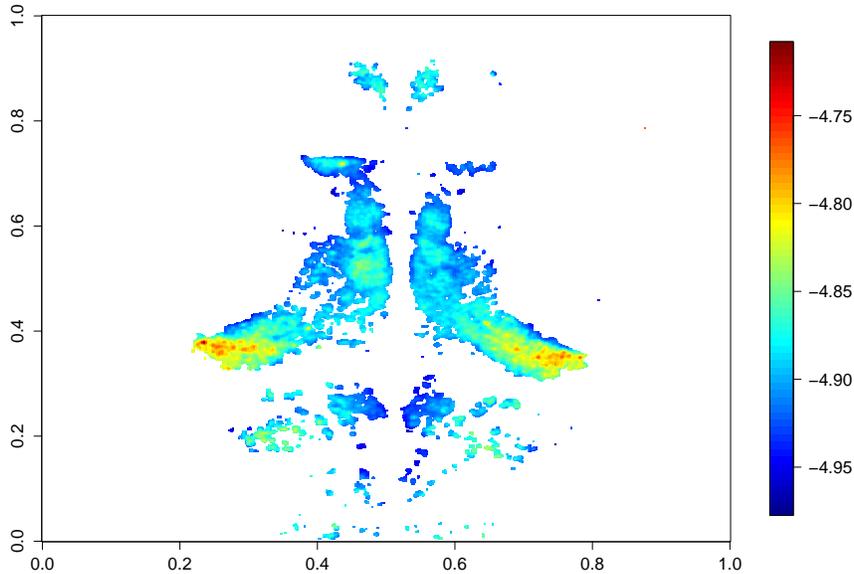

Figure 5: A time-averaged plot of the manually masked zebrafish calcium image. The color of the pixel indicates the time-averaged mean over the 500 time periods.

The pointwise median and 95 percentile interval for each of the 17 clusters are presented in Figure 9.

The segmentation obtained in Figure 8 can be analyzed by scientists to determine the biological significance of the spatial clustering. For example, different clusters may represent different neuronal reaction patterns due to the pharmacologically induced stimulation. Closer analysis of the time-series belonging to each cluster can reveal general deterministic patterns that may lead to an understanding of the nature of such neuronal activities; for example, clusters 15, 16, and 17 appear to have smaller spiking patterns and more volatility along the entire series when compared to the other clusters. A biological discussion is beyond the scope of this article.

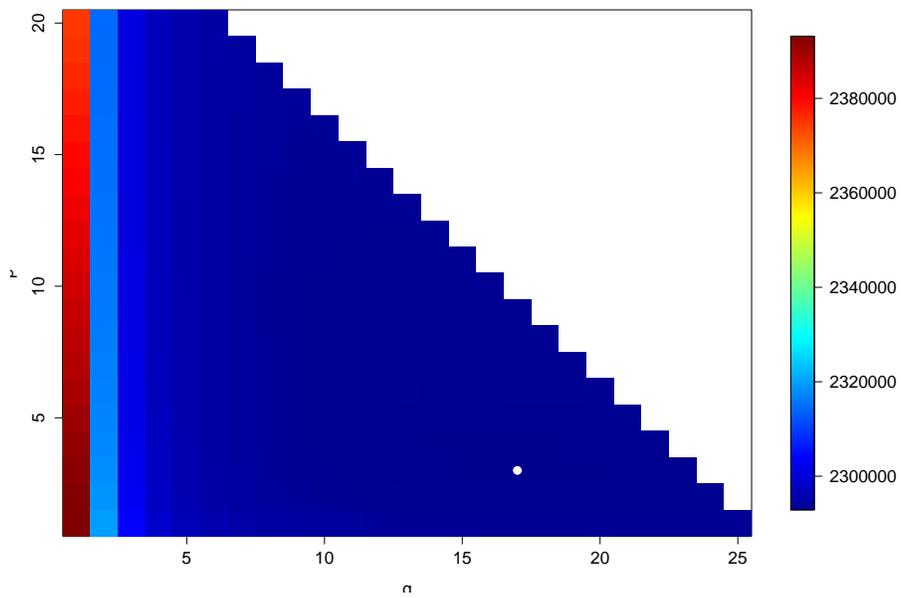

Figure 6: PLIC values for MoAR $(g, p)$ models where $g + p \leq 26$. The white marker indicates the model that minimizes the PLIC value (i.e. $g = 17$ and $p = 3$).

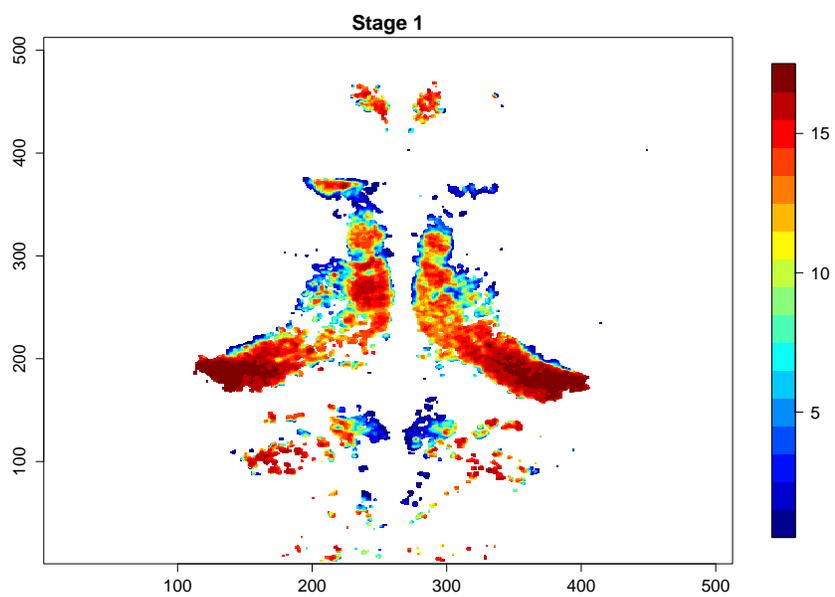

Figure 7: Stage 1 clustering of the manually masked zebrafish calcium image. Each color represents a different cluster.

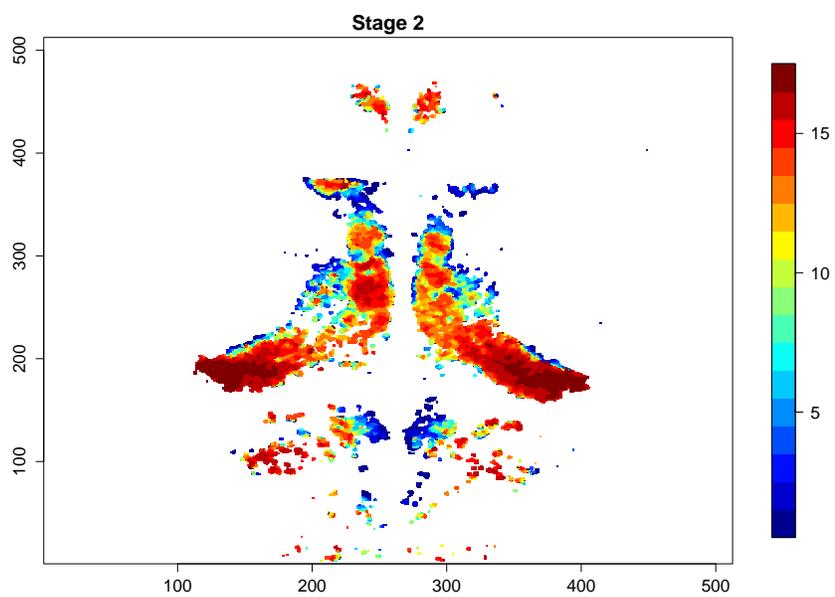

Figure 8: Stage 2 clustering of the manually masked zebrafish calcium image. Each color represents a different cluster.

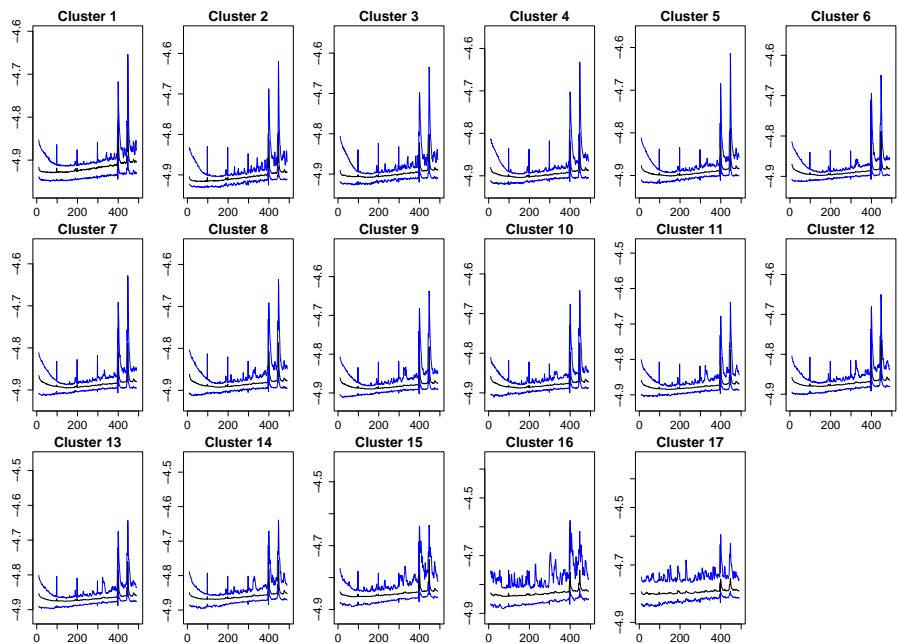

Figure 9: Median and 95 percentile interval for each of clusters that are presented in Figure 8. In each window, the pointwise median is presented in black, and the upper and lower bounds of the pointwise 95 percentile interval are presented in blue.

# 6 Conclusions

In this article, we have introduced a two-stage procedure for the clustering time-series data that arises from a spatially dependent process. In Stage 1 of our methodology, an MoAR model is fitted via MMaL estimation, and Rule (10) is used to marginally cluster the data units. In Stage 2, an MRF is fitted using MPL estimation, and Rule (18), which accounts for the spatial dependencies between the data units.

We show that both the MM algorithms used to perform the MMaL estimation of the MoAR model, and the MPL estimation of the MRF model monotonically increase their respective objectives (i.e. MaL and PL, respectively), in each iteration. Furthermore, both algorithms are shown to be globally convergent. We also show that the MMaL estimator and MPL estimator are both consistent.

Through simulations, we show that our two-stage procedure is highly-capable at clustering spatially correlated time-series data, under situations where mean and mean function based methods would fail. Furthermore, we notice that the addition of Stage 2 greatly increases the concordance between the cluster outcomes and the true classes. The PLIC criterion was demonstrated to be effective for the purpose of selecting the order and number of components of the MoAR models.

Our example analysis shows that our methodology can be applied successfully to the analysis of biological imaging data. We finally note that although our method has been developed for the calcium imaging of zebrafish, there is no boundary to adopting it for the analysis of other time-series images.

# Appendices

## I. Proof of Theorem 2

We follow the notation from Section 2.1. The following lemmas are adapted from Amemiya (1985, Thm. 4.1.2) and Andrews (1992, Thm. 5), respectively.

**Lemma 3.** *Make the following assumptions:*

*(A1) Let $\Theta$ be an open subset of $\mathbb{R}^q$, and let $\boldsymbol{\theta}_0 \in \Theta$.*

*(A2) Let $\ell_{\mathcal{M},n}(\boldsymbol{\theta}) = \ell_{\mathcal{M},n}(\boldsymbol{Y}_1,...,\boldsymbol{Y}_n;\boldsymbol{\theta})$ be a measurable function of $\boldsymbol{Y}_1,...,\boldsymbol{Y}_n$ for all $\boldsymbol{\theta} \in \Theta$, and let $\nabla \ell_{\mathcal{M},n}$ exist and be continuous in some bounded open neighborhood $N_1$ of $\boldsymbol{\theta}_0$.*

*(A3) Let $n^{-1}\ell_{\mathcal{M},n}(\boldsymbol{\theta})$ converge to a nonstochastic function $\ell_{\mathcal{M}}(\boldsymbol{\theta})$ in some bounded open neighborhood $N_2$ of $\boldsymbol{\theta}_0$.*

*(A4) Let $\ell_{\mathcal{M}}(\boldsymbol{\theta})$ attain a strict-local maximum at $\boldsymbol{\theta}_0$.*

*If $\Theta_n = \{\boldsymbol{\theta} : \nabla \ell_{\mathcal{M}} = \mathbf{0}\}$ (where we take $\Theta_n = \{\bar{\boldsymbol{\theta}}\}$, for some $\bar{\boldsymbol{\theta}} \in \Theta$, if $\nabla \ell_{\mathcal{M}} = \mathbf{0}$ has no solution), then for any $\epsilon > 0$,*

$$\lim_{n \to \infty} \mathbb{P}\left[\inf_{\boldsymbol{\theta} \in \Theta_n} (\boldsymbol{\theta} - \boldsymbol{\theta}_0)^T (\boldsymbol{\theta} - \boldsymbol{\theta}_0) > \epsilon\right] = 0.$$

**Lemma 4.** *Make the following assumptions:*

*(B1) Let $\Theta$ be a totally bounded metric space.*

*(B2) Let $n^{-1}\ell_{\mathcal{M},n}(\boldsymbol{\theta}) \xrightarrow{P} \mathbb{E}\log f(\boldsymbol{Y}_s;\boldsymbol{\theta})$, for all $\boldsymbol{\theta} \in \Theta$.*

*(B3) Let $\sup_{\boldsymbol{\theta} \in N_3} |\log f(\boldsymbol{Y}_s;\boldsymbol{\theta})| \leq M$ for some $M < \infty$, for all compact subsets $N_3 \subset \Theta$.*

*If $\log f(\boldsymbol{Y}_s;\boldsymbol{\theta})$ is continuous in $\boldsymbol{\theta}$, uniformly over $\Theta$, and $\boldsymbol{Y}_s$ are identically distributed then,*

$$\sup_{\boldsymbol{\theta} \in N_3} \left|n^{-1}\ell_{\mathcal{M},n}(\boldsymbol{\theta}) - \mathbb{E}\log f(\boldsymbol{Y}_s;\boldsymbol{\theta})\right| \xrightarrow{P} 0.$$

We obtain Theorem 2 by applying Lemma 3. Assumptions A1 and A2 are fulfilled by the definition of the parameter space (i.e. (9)). and by noting that (3) is everywhere smooth, in $\boldsymbol{\theta}$. To validate A3, we require Lemma 4.

Assumption B1 is validated by noting that any bounded subset of $\mathbb{R}^q$ is a totally bounded metric space (note that this implies that we suppose that the parameter components are bounded in absolute value by some large but finite number). We validate B2 by noting that $n^{-1}\ell_{\mathcal{M},n}(\boldsymbol{\theta}) = n^{-1}\sum_{s=1}^{n}\log f(\boldsymbol{Y}_s;\boldsymbol{\theta})$ is a sample average. Since $\boldsymbol{Y}_s$ is ergodic (or mixing), we know that the law of large numbers applies to $\log f(\boldsymbol{Y}_s;\boldsymbol{\theta})$, provided that $|\log f(\boldsymbol{Y}_s;\boldsymbol{\theta})|$ has finite first (and second) moment(s), for fixed $\boldsymbol{\theta}$ (cf. White (2001, Sec. 3.4)). This is easily validated using Atienza et al. (2007, Lem. 1). Since $\log f(\boldsymbol{Y}_s;\boldsymbol{\theta})$ is continuous, B3 is true for any compact set $N_3$, thus we have the conclusion of Lemma 4, which implies A3. Lastly, we require that A4 be made explicitly to obtain the conclusion of Lemma 3, which completes the proof.

## II. Proof of Theorem 4

We follow the notation from Section 3.1. The following lemma is adapted from Geman and Graffigne (1986).

**Lemma 5.** *If $C_1,...,C_n$ is a random sample that is best approximated via a model of form (12) with some parameter $\boldsymbol{\psi}_0$, and (12) is identifiable, then $\hat{\boldsymbol{\psi}}_n \xrightarrow{P} \boldsymbol{\psi}_0$.*

**Proposition 7.** *The MRF (12) is identifiable.*

*Proof.* By Definition 3, we are required to show that if $\boldsymbol{\psi} \neq \boldsymbol{\psi}'$, then

$$\mathbb{P}\left(c_s = i | C_{(s)} = c_{(s)}; \boldsymbol{\psi}\right) \neq \mathbb{P}\left(c_s = i | C_{(s)} = c_{(s)}; \boldsymbol{\psi}'\right),$$

for some $c_s$ and $c_{(s)}$. Note that for any $i \neq i'$, we have

$$\frac{\mathbb{P}\left(c_s = i | C_{(s)} = c_{(s)}; \boldsymbol{\psi}\right)}{\mathbb{P}\left(c_s = i' | C_{(s)} = c_{(s)}; \boldsymbol{\psi}\right)} = \exp\left(\boldsymbol{c}_{is}^T \boldsymbol{\psi}_i - \boldsymbol{c}_{i's}^T \boldsymbol{\psi}_{i'}\right).$$

Thus, if $\boldsymbol{\psi} \neq \boldsymbol{\psi}'$, then $\exp\left(\boldsymbol{c}_{is}^T \boldsymbol{\psi}_i - \boldsymbol{c}_{i's}^T \boldsymbol{\psi}_{i'}\right) = \exp\left(\boldsymbol{c}_{is}^T \boldsymbol{\psi}'_i - \boldsymbol{c}_{i's}^T \boldsymbol{\psi}'_{i'}\right)$, for each $i \neq i'$. This simplifies to the statement, $\boldsymbol{c}_{is}^T \left(\boldsymbol{\psi}_i - \boldsymbol{\psi}'_i\right) + \boldsymbol{c}_{i's}^T \left(\boldsymbol{\psi}'_{i'} - \boldsymbol{\psi}_{i'}\right) = 0$, for all $i \neq i'$, which is only true for all $c_{(s)}$ if $\boldsymbol{\psi} = \boldsymbol{\psi}'$. Thus, we have the result by contradiction. □

Lemma 5 and Proposition (7) together imply Theorem 4.

# References

# References

Akerboom, J., T.-W. Chen, T. J. Wardill, L. Tian, J. S. Marvin, S. Mutlu, N. C. Calderon, F. Esposti, B. G. Borghuis, X. R. Sun, A. Gordus, M. B. Orger, R. Portugues, F. Engert, J. J. Macklin, A. Filosa, A. Aggarwal, R. A. Kerr, R. Takagi, S. Kracun, E. Shigetomi, B. S. Khakh, H. Baier, L. Lagnado, S. S.-H. Wang, C. I. Bargmann, B. E. Kimmel, V. Jayaraman, K. Svoboda, D. S. Kim, E. R. Schreiter, and L. L. Looger (2012). Optimization of a GCaMP calcium indicator for neural activity imaging. *Journal of Neuroscience 32*, 13819–13840.

Amemiya, T. (1985). *Advanced Econometrics*. Cambridge: Harvard University Press.

Andrews, D. W. K. (1992). Generic uniform convergence. *Econometric Theory 8*, 241–257.

Atienza, N., J. Garcia-Heras, J. M. Munoz-Pichardo, and R. Villa (2007). On the consistency of MLE in finite mixture models of exponential families. *Journal of Statistical Planning and Inference 137*, 496–505.

Box, G. E. P., G. M. Jenkins, and G. C. Reinsel (2008). *Time Series Analysis* (4 ed.). New York: Wiley.

Boyd, S. and L. Vandenberghe (2004). *Convex Optimization*. Cambridge: Cambridge University Press.

Cadez, I. V., S. Gaffney, and P. Smyth (2000, August). A general probabilistic framework for clustering individuals and objects. In *Proceedings of the Sixth ACM SIGKDD International Conference on Knowledge Discovery and Data Mining*, Boston MA, USA, pp. 140–149.

Celeux, G., O. Martin, and C. Lavergne (2005). Mixture of linear mixed models for clustering gene expression profiles from repeated microarray experiments. *Statistical Modelling 5*, 243–267.

DeSarbo, W. S. and W. L. Cron (1988). A maximum likelihood methodology for clusterwise linear regressions. *Journal of Classification 5*, 249–282.

Eddelbuettel, D. (2013). *Seamless R and C++ Integration with Rcpp*. New York: Springer.

Esling, P. and C. Agon (2012). Time-series data mining. *ACM Computing Surveys 45*, 12:1–12:34.

Fraley, C. and A. E. Raftery (2002). Model-based clustering, discriminant analysis, and density estimation. *Journal of the American Statistical Association 97*, 611–631.

Fraley, C. and A. E. Raftery (2003). Enhanced model-based clustering, density estimation, and discriminant analysis software: MCLUST. *Journal of Classification 20*, 263–286.

Geman, S. and C. Graffigne (1986). Markov random field image models and their applications to computer vision. In *Proceedings of the International Congress of Mathematicians*, pp. 1496–1517.

Hartvig, N. V. and J. L. Jensen (2000). Spatial mixture modeling of fMRI data. *Human Brain Mapping 11*, 233–248.

Hubert, L. and P. Arabie (1985). Comparing partitions. *Journal of Classification 2*, 193–218.

Ibragimov, I. A. and Y. V. Linnik (1971). *Independent and Stationary Sequences of Random Variables*. Groningen: Wolters-Noordhoff.

Ji, C. and L. Seymour (1996). A consistent model selection procedure for Markov random fields based on penalized pseudolikelihood. *Annals of Applied Probability 6*, 423–443.

Jones, P. N. and G. J. McLachlan (1992). Fitting finite mixture models in a regression context. *Australian Journal of Statistics 34*, 233–240.

Lange, K. (2013). *Optimization*. New York: Springer.

Liao, T. W. (2005). Clustering of time series data—a survey. *Pattern Recognition 38*, 1857–1874.

Luan, Y. and H. Li (2003). Clustering of time-course gene expression data using a mixed-effects model with B-splines. *Bioinformatics 19*, 474–482.

MacQueen, J. (1967). Some methods for classification and analysis of multivariate observations. In *Proceedings of the fifth Berkley symposium on mathematical statistics and probability*.

McLachlan, G. J. (1992). *Discriminant Analysis And Statistical Pattern Recognition*. New York: Wiley.

McLachlan, G. J. and K. E. Basford (1988). *Mixture Models: Inference And Applications To Clustering*. New York: Marcel Dekker.

McLachlan, G. J. and T. Krishnan (2008). *The EM Algorithm And Extensions* (2 ed.). New York: Wiley.

McLachlan, G. J. and D. Peel (2000). *Finite Mixture Models*. New York: Wiley.

McLachlan, G. J., D. Peel, K. E. Basford, and P. Adams (1999). The EMMIX software for the fitting of mixtures of normal and t-components. *Journal of Statistical Software 4*, 1–14.

Muto, A. and K. Kawakami (2013). Prey capture in zebrafish larvae serves as a model to study cognitive functions. *Frontiers in Neural Circuits 7*, 1–5.

Ng, S. K., G. J. McLachlan, K. W. L. Ben-Tovim, and S. W. Ng (2006). A mixture model with random-effects components for clustering correlated gene-expression profiles. *Bioinformatics 22*, 1745–1752.

Nguyen, H. D. and G. J. McLachlan (2015). Maximum likelihood estimation of Gaussian mixture models without matrix operations. *Advances in Data Analysis and Classification 9*, 371–394.

Nguyen, H. D., G. J. McLachlan, N. Cherbuin, and A. L. Janke (2014). False discovery rate control in magnetic resonance imaging studies via Markov random fields. *IEEE Transactions on Medical Imaging 33*, 1735–1748.

R Core Team (2013). *R: a language and environment for statistical computing*. R Foundation for Statistical Computing.

Razaviyayn, M., M. Hong, and Z.-Q. Luo (2013). A unified convergence analysis of block successive minimization methods for nonsmooth optimization. *SIAM Journal of Optimization 23*, 1126–1153.

Same, A., F. Chamroukhi, G. Govaert, and P. Aknin (2011). Model-based clustering and segmentation of time series with change in regime. *Advances in Data Analysis and Classification 5*, 301–321.

Scharl, T., B. Grun, and F. Leisch (2010). Mixtures of regression models for time course gene expression data: evalluation of initialization and random effects. *Bioinformatics 26*, 370–377.

Schwarz, G. (1978). Estimating the dimensions of a model. *Annals of Statistics 6*, 461–464.

Stanford, C. D. and A. E. Raftery (2002). Approximate Bayes factor for image segmentation: the pseudolikelihood information criterion (PLIC). *IEEE Transactions on Pattern Analysis and Machine Intelligence 24*, 1517–1520.

Titterington, D. M., A. F. M. Smith, and U. E. Makov (1985). *Statistical Analysis Of Finite Mixture Distributions*. New York: Wiley.

Varin, C. (2008). On composite marginal likelihoods. *Advances in Statistical Analysis 92*, 1–28.

Vincent, T., L. Risser, and P. Ciuciu (2010). Spatially adaptive mixture modeling for analysis of fMRI time series. *IEEE Transactions on Medical Imaging 29*, 1059–1074.

Wang, K., S. K. Ng, and G. J. McLachlan (2012). Clustering of time-course gene expression profiles using normal mixed models with autoregressive random effects. *BMC Bioinformatics 13*, 300.

White, H. (2001). *Asymptotic Theory For Econometricians*. San Diego: Academic Press.

Woolrich, M. W., T. E. J. Behrens, C. F. Beckmann, and S. M. Smith (2005). Mixture models with adaptive spatial regulariaztion for segmentation with an application to FMRI data. *IEEE Transactions on Medical Imaging 24*, 1–11.

Xiong, Y. and D.-Y. Yeung (2004). Time series clustering with ARMA mixtures. *Pattern Recognition 37*, 1675–1689.